
\input phyzzx.tex
\input pictex.tex
\date{November 1991}
\hfill{CLNS 91/1117}
\titlepage
\title{\bf Restricted Sine-Gordon Theory in the Repulsive Regime as
Perturbed Minimal CFTs}
\author{Changrim Ahn
\foot{E-mail address: AHN@CORNELLA.bitnet}}
\address{\rm F. R. Newman Lab. of Nuclear Studies\break
Cornell University\break
	  Ithaca, NY 14853-5001}

\abstract

We construct the restricted sine-Gordon theory by truncating the
sine-Gordon multi-soliton Hilbert space for the repulsive coupling
constant due to
the quantum group symmetry $SL_q(2)$ which we identify from
the Korepin's $S$-matrices.
We connect this restricted sine-Gordon theory with
the minimal ($c<1$) conformal field theory ${\cal M}_{p/p+2}$ ($p$ odd)
perturbed by the least relevent primary field $\Phi_{1,3}$.
The exact $S$-matrices are derived for the particle spectrum of a kink
and neutral particles.
As a consistency check, we compute the
central charge of the restricted theory in the UV limit
using the thermodynamic Bethe ansatz analysis and show that it is equal
to
that of ${\cal M}_{p/p+2}$.

\endpage
\def\semi{;\hfill\break}
\REF\BPZ{A. A. Belavin, A. M. Polyakov, and A. B. Zamolodchikov,
Nucl. Phys. {\bf B241} (1984) 333.}
\REF\zami{A. B. Zamolodchikov, Int. Journ. of Mod. Phys. A4 (1989)
4235.}
\REF\ABF{G. E. Andrew, R. J. Baxter, and P. J. Forrester, J. Stat. Phys.
{\bf 35} (1984) 193.}
\REF\RSG{A. LeClair, Phys. Lett. {\bf 230B} (1989) 103\semi
	D. Bernard and A. LeClair, Nucl. Phys. {\bf B340} (1990) 721.}
\REF\Smirnov{F. Smirnov, Int. J. Mod. Phys. {\bf A4} (1989) 4213.}
\REF\RaS{N. Yu Reshetikhin and F. Smirnov, Comm. Math. Phys. {\bf 131}
(1990) 157.}
\REF\BaL{D. Bernard and A. LeClair, Phys. Lett. {\bf B247} (1990) 309.}
\REF\EaY{T. Eguchi and S-K. Yang, Phys. Lett. {\bf B224} (1989) 373;
		{\bf B235} (1990) 282.}
\REF\ABL{C. Ahn, D. Bernard, and A. LeClair, Nucl. Phys. {\bf B336}
(1990) 409.}
\REF\Ahn{C. Ahn, Nucl. Phys. {\bf B354} (1991) 57.}
\REF\smirii{F. A. Smirnov, Int. J. Mod. Phys. {\bf A6} (1991) 1407.}
\REF\Felder{G. Felder, {\it Nucl. Phys.} {\bf B317} (1989) 215.}
\REF\FQS{D. Friedan, Z. Qiu, and S. Shenker, Phys. Rev. Lett {\bf 52}
(1984) 1575.}
\REF\Coleman{S. Coleman, Phys. Rev. {\bf D11} (1975) 2088.}
\REF\DHN{F. Dashen, B. Hasslacher, and A. Neveu, Phys. Rev. {\bf D11}
(1975) 3424.}
\REF\FaK{V. E. Korepin and L. D. Faddeev, Teor. Mat. Fiz. {\bf 25}
(1975) 147.}
\REF\Zamol{A. B. Zamolodchikov, Comm. Math. Phys. {\bf 55}
(1977) 183\semi
A. B. Zamolodchikov and Al. B. Zamolodchikov, Annals.
Phys. 120 (1979)  253.}
\REF\KaT{M. Karowski and H. J. Thun, Nucl. Phys. B130 (1977)  295.}
\REF\BaT{H. Bergknoff and H. B. Thacker, Phys. Rev. Lett. {\bf 42}
(1979) 135; Phys. Rev. {\bf D19} (1979) 3666.}
\REF\Korepini{V. E. Korepin, Teor. Mat. Fiz. {\bf 41} (1979) 169.}
\REF\Korepin{V. E. Korepin, Comm. Math. Phys. {\bf 76} (1980) 165.}
\REF\BKI{N. M. Bogoliubov, V. E. Korepin, and A. G. Izergin,
Phys. Lett. {\bf B159} (1985) 345.}
\REF\JNW{G. I. Japaridze, A. A. Nersesyan, and P. B. Wiegmann,
Physica Scripta {\bf 27} (1983) 5.}
\REF\KaR{A. N. Kirillov and N. Reshetikhin, LOMI preprint,
`Representations of the Algebra ${\cal U}_q[sl(2)]$, $q$-orthogonal
Polynomials and Invariants of Links'.}
\REF\TBA{Al. B. Zamolodchikov, Nucl. Phys. {\bf B342} (1990) 695\semi
T. Klassen and E. Melzer, Nucl. Phys. {\bf B338} (1990) 485;
{\bf B350} (1991) 635\semi
M. J. Martins, Phys. Lett. {\bf 240B} (1990) 404\semi
P. Christe and M. J. Martins, Mod. Phys. Lett. {\bf A5} (1990) 2189.}
\REF\IaM{H. Itoyama and P. Moxhay, Phys. Rev. Lett. {\bf 65} (1990)
2102.}
\REF\zamolii{Al. B. Zamolodchikov, Nucl. Phys. {\bf B358} (1991) 497;
Preprint ENS-LPS-331 (1991).}
\REF\AaN{C. Ahn and S. Nam, CLNS 91/1095, SNUTP-91-30, to appear in
Phys. Lett. {\bf B}.}

\chapter{Introduction}

Much progress
have been made recently on integrable quantum field theories (QFTs)
from the conformally invariant 2D field theories\refmark\BPZ\ (CFTs)
perturbed by the relevent operators.\refmark\zami\
For certain CFTs, the integrability
can be preserved under massive perturbations which breaks the conformal
symmetry.
The resulting integrable QFTs have been identified with various
well-known integrable models as well as quite new ones.

There have been two major approaches on the integrable perturbations of
CFTs. One is a direct method of constructing  infinite number of
conserved currents explicitly.
After constructing the integrability, one derives the particle spectrum and
exact $S$-matrices using the bootstrap conjecture.\refmark\zami\
This method, however, has a limitation on its applicability
in that it is not easy to construct
the infinite number of conserved currents explicitly for general classes
of  the CFTs and perturbations.

The other approach is to construct new integrable field theories out of
known one and identify the new theories with perturbed CFTs.
New theories are constructed by restricting Hilbert space of the
starting integrable QFT in such a manner that maintains the integrability.
This is an QFT analogy of the `restricted solid-on-solid'
(RSOS) model in the integrable lattice model.\refmark\ABF\
Many new integrable QFTs with exact $S$-matrices have been obtained by this
method and identified with perturbed CFTs.\refmark{\RSG-\Ahn}\
Most of these new integrable QFTs are obtained from
the sine-Gordon (SG) theory as the
starting integrable QFT.
The restricted sine-Gordon (RSG) theory has been constructed and
identified with minimal $(c<1)$ CFTs perturbed by the least relevent
operator.\refmark{\RSG-\RaS}\
This has been extended to the general coset CFTs perturbed by the
least relevent operator and connected with many integrable QFTs with
extra symmetries, such as supersymmetry.\refmark{\ABL,\Ahn}\
Recently, new restricted integrable QFTs have been constructed from
Zhiber-Mikhailov-Shabat model ($A^{(2)}_2$ affine Toda model with
imaginary coupling constant) and
related to the minimal CFTs perturbed by $\Phi_{1,2}$ and $\Phi_{2,1}$
operators.\refmark\smirii\

In this paper, we continue this approach and extend the RSG theory
to a different
region of the SG coupling constant, the repulsive regime.
All the previous results on the RSG theory \refmark{\RSG-\Ahn}\
were based on the SG theory in the attractive regime and special
points in the repulsive regime.
Outside of this region,
the SG theory behaves quite differently.
We will construct new RSG theory (particle spectrum and $S$-matrices)
for this new region
and identify this theory with minimal CFT ($c<1$) ${\cal M}_{p/p+2}$
($p$ odd) perturbed by the least relevent field $\Phi_{1,3}$.

The RSG theory can be constructed in a similar way as
minimal CFTs through the Feigin-Fuchs-Felder method.
One can think of the SG theory as a massive perturbation
of $c=1$ free boson theory by the periodic potential.
After introducing a background charge, one restricts the
Fock space of the boson using the underlying BRST cohomology structure
to get $c<1$ minimal CFTs ${\cal M}_{p/q}$.\refmark\Felder\
The analogy of this procedure in the massive theory
is the restriction of the multi-soliton
Hilbert space based on the quantum group symmetry of the SG theory.
The resulting theory, the RSG
theory, is identified with the minimal CFTs ${\cal M}_{p/q}$ perturbed
by $\Phi_{1,3}$.
In a similar way as unitarity condition on the physical state further reduces
the minimal CFTs to unitary CFTs ${\cal M}_{p/p+1}$,\refmark\FQS\
the unitarity condition on the $S$-matrices of the RSG theory
allowes only a few inifinite series out of the minimal CFTs to yield
the unitary $S$-matrices after the massive perturbation.\refmark\RaS\

This paper is organized as follows:
In sect.2, we review the major relevent results on the SG theory
in the historical order.
We will treat the SG theory and the massive Thirring model (MTM) as
equivalent theories following S. Coleman's discovery.\refmark\Coleman\
In terms of the MTM coupling constant, we classify the SG (MTM) theory
into two regimes, attractive and repulsive regimes.
We present semi-classical results on the bound state mass spectrum
in the attractive regime,\refmark\DHN\ from which A.B.
Zamolodchikov conjectured the familiar exact soliton $S$-matrix.
\refmark\Zamol\
The spectrum consists of the solitons, anti-solitons, and their bound
states (the breathers). This result in the attractive regime
has been confirmed by the Bethe ansatz.\refmark{\BaT,\Korepini}\

The situation in the repulsive regime
\foot{By the repulsive regime, we mean the region except discrete
first-order phase transition points.}
is quite different.
The physical vacuum constructed in the attractive regime turns out to be
not true vacuum. It is V. Korepin who first constructed the true vacuum
to get the physical spectrum of the MTM theory.\refmark\Korepin\
The spectrum in this case consists of the solitons, anti-solitons, and
neutral particles which are not bound states. Exact $S$-matrices of
these particles have been also derived. Our construction of the RSG
theory in the repulsive regime will be based on this result.

We explain the RSG theory in sect.3
starting with the $SL_q(2)$ quantum group symmetry in the SG theory.
\refmark{\RSG-\EaY}\
We derive the $SL_q(2)$ symmetry of the SG theory in the repulsive
regime where the deformation parameter $q$ becomes modified. Using this
quantum group symmetry, we can restrict the multi-soliton Hilbert space
to obtain the RSG theory in the repulsive regime. After imposing the
unitarity condition on the $S$-matrices, we identify the RSG theory with
perturbed minimal CFTs. The particle spectrum includes a kink and many
neutral particles with all different masses.

In sect.4, we check these $S$-matrices of the new
RSG theory using the thermodynamic
Bethe ansatz.\refmark{\TBA-\AaN}\
This proves the consistency of our construction of the new RSG theory
as the minimal CFTs ${\cal M}_{p/p+2}$
perturbed by the $\Phi_{1,3}$ operator.
We will conclude with a few comments and open questions in sect.5.


\def\t{\theta}
\def\g{\gamma}
\chapter{$S$-matrices of the sine-Gordon Theory}

In this section, we review the results on the SG theory which will be
used in later sections.
In particular, we explain the Bethe ansatz which gives the exact results in
the both attractive and repulsive regime.

\section{the sine-gordon theory}

We start with the SG Lagrangian
$$ {\cal L}_{\rm SG}= {1\over{2}} (\partial_{\mu}\phi)^2
	  + {M^2_0\over{\beta^2}} \cos(\beta\phi),\eqn\sglag$$
where $M_0$ is a mass scale and $\beta$ is
a coupling constant.
Due to the periodic potential, the SG theory has classical soliton (and
anti-soliton) solutions which connect two different vacua at
$x=\pm\infty$. Each solitonic solution carries a
conserved topological charge.

S. Coleman discovered\refmark\Coleman\
that the quantum SG theory is equivalent to
the massive Thirring model
$$\eqalign{{\cal L}_{\rm MTM}
&= i{\overline \psi}\gamma_{\mu}\partial_{\mu}\psi
	   -m_0{\overline \psi}\psi -{g\over{2}}({\overline \psi}\gamma_{\mu}
	   \psi)^2,\cr
{g\over{\pi}}&={4\pi\over{\beta^2}}-1,\quad{\rm for }
\quad 0\le {\beta^2\over{8\pi}}<1,\cr}\eqn\mtmlag$$
and identified the SG soliton with the Thirring fermion.
Therefore, we can consider the SG theory and MTM as two different
realizations of the same theory.

He also showed that if
$\beta^2/8\pi\ge 1$ the energy density of the theory is
unbounded from below and not well-defined.
Throughout this paper, we will consider only the case of $\beta^2/8\pi< 1$.
We define the following parameters for the later convenience:
$$P={\gamma\over{8\pi}}={\beta^2/8\pi\over{1-\beta^2/8\pi}},\qquad
{\rm and}\qquad {\mu\over{\pi}}=1-{\beta^2\over{8\pi}}.\eqn\coupling$$
The Thirring interaction becomes
$$\eqalign{
&{\rm attractive}\ (g>0)\quad{\rm for}\quad \beta^2<4\pi,
\quad {\g\over{8\pi}}<1,\quad \mu>\pi/2,\cr
&{\rm repulsive}\ (g<0)\quad{\rm for}\quad \beta^2>4\pi,
\quad {\g\over{8\pi}}>1,\quad \mu<\pi/2.\cr}\eqn\regime$$

The complete particle spectrum of the SG (MTM) theory was derived in the
semi-classical (WKB) approximation and claimed to be exact even in the
full quantum theory.\refmark\DHN\
In the attractive regime ($\gamma /8\pi<1$), the spectrum consists of
the SG soliton ($A^+$), anti-soliton ($A^-$) with the mass $m_f$ and
the bound states of these, the breathers ($B_n$), with  the masses
$$m_n=2m_f\sin\left({n\g\over{16}}\right);\quad n=1,2,\cdots<{8\pi\over{\g}}.
								\eqn\massi$$
The first exact $S$-matrix of the SG (MTM) theory was conjectured by
Faddeev and Korepin\refmark\FaK\
for the special values in the attractive regime $\g/8\pi=1/n$
with $n=1,2,\cdots$ (the soliton sector),
$$S^{SG}_{++,++}\left(\t,{\g\over{8\pi}}\right)
=e^{in\pi}\prod^n_{k=1}\left[{\exp(\t-i\pi k/n)+1
	 \over{\exp{\t}+\exp(-i\pi k/n)}}\right].\eqn\fadeev$$

\section{Zamolodchikov's $S$-matrices}

The SG (MTM) theory is integrable in
the sense that one can construct infinite number of commuting conserved
currents in the both classical and quantum levels.
In the quantum theory, this means that
the $S$-matrices are elastic without any particle creation or
annihilation. The multi-particle $S$-matrices are completely factorized
into the products of two-particle $S$-matrices.
These two-particle $S$-matrices should satisfy Yang-Baxter equations
which determine the $S$-matrices completely along with unitarity and
crossing symmetry upto the overall CDD factor. We will consider
only the case where additional CDD factors are set to one (minimal
solution).

A.B. Zamolodchikov derived the first exact
$S$-matrix of the SG solitons and anti-solitons,
assuming that the semi-classical masses in
Eq.\massi\ are exact. This means that the soliton $S$-matrix should
have poles in the physical strip which corespond to the bound states.
Along with the unitarity,
crossing symmetry,
and the known exact results for the special attractive couplings
in Eq.\fadeev, the SG soliton $S$-matrix can be decided
completely in the attractive regime.\refmark\Zamol\
The $S$-matrix of the SG soliton ($A^+$) and anti-soliton ($A^-$)
(or the Thirring fermion and antifermion) is equal to
$$\eqalign{S^{SG}_{++,++}\left(\t,{\g\over{8\pi}}\right)&=
S^{SG}_{--,--}\left(\t,{\g\over{8\pi}}\right)={U(\t)\over{i\pi}}
\sinh\left[{8\pi\over{\g}}(i\pi-\t)\right]  \cr
S^{SG}\left(\t,{\g\over{8\pi}}\right)&= {U(\t)\over{i\pi}}\left\{
\bordermatrix{&A^+A^-&A^-A^+\cr A^+A^-&\sinh{8\pi\t\over{\g}}
&i\sin{8\pi^2\over{\g}}\cr A^-A^+&i\sin{8\pi^2\over{\g}}&
\sinh{8\pi\t\over{\g}} \cr }\right\},  \cr}   \eqn\sgsma$$
with $U(\t)$, satisfying $U(\t)=U(i\pi-\t)$,
$$\eqalign{  U(\t)&=\Gamma\left({8\pi\over{\g}}\right)\Gamma\left(1+i
{8\t\over{\g}}\right)\Gamma\left(1-{8\pi\over{\g}}-i{8\t\over{\g}}\right)
\prod^{\infty}_{n=1}{R_n(\t)R_n(i\pi-\t)\over{R_n(0)R_n(i\pi)}},\cr
R_n(\t)&={\Gamma\left(2n{8\pi\over{\g}}+i{8\t\over{\g}}\right)
\Gamma\left(1+2n{8\pi\over{\g}}+i{8\t\over{\g}}\right)  \over{
\Gamma\left((2n+1){8\pi\over{\g}}+i{8\t\over{\g}}\right)
\Gamma\left(1+(2n-1){8\pi\over{\g}}+i{8\t\over{\g}}\right) }}.
								\cr}\eqn\utheta$$
One can express these with an integral representation,
$$S^{SG}_{++,++}\left(\t,{\g\over{8\pi}}\right)
=\exp\left[-\int_0^{\infty}{dx\over{x}}
{\sinh((4\pi x/\g)-x/2)\sinh(8i\t x/\g)\over{
\sinh(x/2)\cosh(4\pi x/\g)}}\right].\eqn\integ$$
The soliton $S$-matrix \sgsma\ has been checked perturbatively based on
the MTM.

The $S$-matrices between the breathers and solitons
have been derived from
three-solitons (and anti-solitons)
scattering processes,\refmark\KaT\
$$S_{\rm SG}^{(n)}\left(\t,{\g\over{8\pi}}\right)={\sinh\t+i\cos{n\g\over{16}}
\over{\sinh\t-i\cos{n\g\over{16}}}}
\ \prod_{l=1}^{n-1}{\sin^2\left({n-2l\over{32}}\g-{\pi\over{4}}+
i{\t\over{2}}\right)\over{ \sin^2\left({n-2l\over{32}}\g-{\pi\over{4}}-
i{\t\over{2}}\right)} },\eqn\sgn$$
for the process $A^{\pm}+B_n\rightarrow A^{\pm}+B_n$.
The $S$-matrices between the breathers (four (anti) solitons process) are
equal to\refmark\KaT\
$$\eqalign{
S_{\rm SG}^{(n,m)}&\left(\t,{\g\over{8\pi}}\right)
={\sinh\theta +i\sin\left({n+m\over{16}}\g\right)\over
{\sinh\theta -i\sin\left({n+m\over{16}}\g\right)}}
{\sinh\theta +i\sin\left({n-m\over{16}}\g\right)\over
{\sinh\theta -i\sin\left({n-m\over{16}}\g\right)}}\cr
&\quad\times \prod_{l=1}^{{\rm min}(m,n)-1}
{\sin^2\left({m-n-2l\over{32}}\g+i{\t\over{2}}\right)
\cos^2\left({m+n-2l\over{32}}\g+i{\t\over{2}}\right)
\over{\sin^2\left({m-n-2l\over{32}}\g+i{\t\over{2}}\right)
\cos^2\left({m+n-2l\over{32}}\g+i{\t\over{2}}\right)}},\cr}\eqn\sgmn$$
for the process $B_n(\t_1)+B_m(\t_2)\rightarrow B_n(\t_1)+B_m(\t_2)$
with $n\ge m$.
One can check these $S$-matrices by comparing Eq.\sgmn\ for $m=n=1$ with
the perturbative computation based on the SG Lagrangian \sglag\
because one can identify the lowest massive breather $B_1$ with the
fundamental field $\phi$ in the SG Lagrangian.

We want to emphasize that
Eqs.\sgsma, \sgn\ and \sgmn\ are
based on the semi-classical mass spectrum and
scattering amplitudes \fadeev\ in the attractive regime.
Therefore, these results are valid in principle only in this regime.
In the next subsection, we show that the Zamolodchikov's results are
exact in the attractive regime while new results are obtained in the
repulsive regime using the Bethe ansatz method.

\section{Bethe Ansatz Approach}

The MTM theory has been solved exactly by
the Bethe ansatz. The exact mass spectrum
\refmark\BaT\ and $S$-matrices\refmark\Korepini\ have
been obtained in the attractive regime, which are consistent with the
Zamolodchikov's results. Furthermore, this method has been extended to
the repulsive regime\refmark\Korepin\
to get the exact results which are quite different
from the previous ones.
\foot{The Bethe ansatz solution of the SG theory regularized on the
lattice has given consistent results with these.\refmark\BKI}

The second quantized Hamiltonian of the MTM is from Eq.\mtmlag,
$${\widehat H}=\int dx\left[-i\left(\psi_1^{\dagger}
{\partial\over{\partial x}}\psi_1-\psi_2^{\dagger}{\partial\over{\partial
x}}\psi_2\right) + m_0(\psi_1^{\dagger}\psi_2+\psi_2^{\dagger}\psi_1)
+2g\psi_1^{\dagger}\psi_2^{\dagger}\psi_2\psi_1\right].\eqn\mtmham$$
Note that due to the negative sign of the second term in the kinetic energy,
there are negative energy modes. This means that the vacuum $|0\rangle$,
defined by $\psi_{1,2}(x)|0\rangle=0$, and states constructed
upon it are not true physical states. We refer to
these states created by the fermion modes acting on the false vacuum
$|0\rangle$ as `pseudo-particles'.
The true physical states and particles should be constructed out of the
true vacuum which is a Dirac sea where all the negative
energy states are completely filled.

One can diagonalize the Hamiltonian with the Bethe ansatz wave function
$$\eqalign{|\Phi&(\t_1,\cdots,\t_N)\rangle= \int\prod_{i=1}^N
(e^{im_0 x_i\sinh\t_i}dx_i)\cr
&\times\prod_{i<j\le N}\exp\left[{i\over{2}}\epsilon(x_j-x_i)\phi
(\t_j-\t_i)\right]
A^{\dagger}(\t_1,x_1)\cdots A^{\dagger}(\t_N,x_N)|0\rangle,\cr}
						  \eqn\bethe$$
where $\epsilon(x)$ is the sign-function and $\phi(\t)$ is the phase of the
scattering amplitude $S(\t)$ of the pseudo-particles,
$$S(\t)=\exp[i\phi(\t)]=-{\sinh{1\over{2}}(\t-2i\mu)
\over{\sinh{1\over{2}}(\t+2i\mu)}},\eqn\phase$$
in terms of the parameter $\mu$ defined in Eq.\coupling.
By applying the Hamiltonian \mtmham\ to the state \bethe, one can find
$${\widehat H}|\Phi(\t_1,\cdots,\t_N)\rangle=\left(\sum_{i=1}^N
m_0\cosh\t_i\right)|\Phi(\t_1,\cdots,\t_N)\rangle.
\eqn\energy$$
The operator $A^{\dagger}(\t,x)$ is  related to the fermion operators
by
$$A^{\dagger}(\t,x)=(2\cos\t)^{-1/2}\left[e^{\t/2}\psi_1^{\dagger}(x)
+e^{-\t/2}\psi_2^{\dagger}(x)\right],\eqn\mode$$
where the rapidity $\t$ has complex values in general.
In rapidity space, the Fourier transformed operator
$$A^{\dagger}(\t)=\int
dxe^{-ixm_0\sinh\t}A^{\dagger}(\t,x),\eqn\modeii$$
creates positive energy modes for $\t=\alpha$ and negative energy
modes for $\t=\alpha+i\pi$.
We use $\alpha$ for a real value of the rapidity.
The physical vacuum is the Dirac sea where all the rapidity states
$\t=\alpha+i\pi$ are completely filled.

The physical spectrum can be determined by imposing
two conditions on the Bethe wave function \bethe.
The first one is that  the wave function (the
integrand in Eq.\bethe) should
go to zero as $x_i\rightarrow\pm\infty$.
If the rapidities have imaginary parts, Eq.\bethe\ diverges unless
the scattering amplitudes $S(\t_i-\t_j)$ vanish.
This requires that $\t_i-\t_{i+1}=2i(\pi-\mu)$ from Eq.\phase.
With the condition that the imaginary parts of all the allowed
rapidities lie between $-i\pi$ and $i\pi$,
the only allowed excited rapidity states are `string states'
($S_n(\alpha)$ with $n=1,2,\cdots$) which are composed of $n$
pseudo-particles with the rapidities (in the attractive regime)
$$\eqalign{\t_l&=\alpha+i(\pi-\mu)(n-1-2l),\quad{\rm for}\quad
{\pi\over{2}}<\mu<\pi,\cr
l&=0,1,\cdots,n-1\quad{\rm with}\quad (\pi-\mu)(n-1)<\pi.\cr}\eqn\stringi$$

The other condition is the periodic boundary condition (PBC) in the large
box with length $L$.
{}From PBC on Eq.\bethe ($x_i\rightarrow x_i +L$),
one can find the following equation for each $i$:
$$e^{im_0 L\sinh\t_i}\prod_{i\neq
j}\exp\left[i\phi(\t_j-\t_i)\right]=1.
				 \eqn\period$$
Taking logarithms on both sides, one gets
$$-m_0 L\sinh\t_i=\sum_{i\neq j}\phi(\t_j-\t_i)
+2\pi n_i,\eqn\transcen$$
with arbitrary integers $n_i$.
As $L\to\infty$, in terms of the
density of the rapidity states per unit length,
$$\rho(\t)=\lim_{L\to\infty}{1\over{L(\t_j-\t_{j+1})}},\eqn\den$$
Eq.\transcen\ becomes an integral equation for the
density of the rapidity states in the Dirac sea at $\t=\alpha+i\pi$:
$$2\pi\rho(\t)=-m_0\cosh\t-\int_{-\Lambda+i\pi}^{\Lambda+i\pi}
d\t^{\prime}\left[{\partial\over{\partial\t}}\phi(\t-\t^{\prime})\right]\rho
(\t^{\prime}).\eqn\inteq$$
Note that the density of the allowed states in the left-hand side
(the density of
$n_i$) are the same as that of the actual occupied states $\rho(\t)$
because all the rapidity states in the vacuum (the
Dirac sea) are completely filled by definition.
We introduced the rapidity cut-off parameter $\Lambda$ to
regularize the UV divergence. The cut-off dependence will be absorbed
into the mass renormalization.

\def\rhoh{{\widehat\rho}}
The physical excited states are the string states $S_n(\alpha)$
made of $n$ pseudo-particles from the Dirac sea.
Therefore, an physical excited state consists of a string state $S_n$
and $n$ `holes' in the Dirac sea.
Due to these holes, the pseudo-particle rapidity states in the Dirac sea
are rearranged, which are determined by the following PBC equation:
$$\eqalign{2\pi\rhoh(\t)=&-m_0\cosh\t-\sum_{l=0}^{n-1}
\phi(\t-\t_l)-n\phi(\t-\t^{(h)})\cr
&-\int_{-\Lambda+i\pi}^{\Lambda+i\pi}
d\t^{\prime}\left[{\partial\over{\partial\t}}\phi(\t-\t^{\prime})\right]\rhoh
(\t^{\prime}),\cr}\eqn\inteqi$$
with $\t^{(h)}=\alpha_s+i\pi$ and $\t_l$ in Eq.\stringi.
The energy of this excited state relative to the vacuum is equal to
$$\eqalign{E_n&=m_n\cos\alpha\cr
&=\sum_{l=0}^{n-1}m_0\cosh\t_l-nm_0\cosh\t^{(h)}
	 +m_0\int_{-\Lambda+i\pi}^{\Lambda+i\pi}d\t\left[\rho(\t)
-\rhoh(\t)\right]\cosh\t.\cr}\eqn\exenergy$$

In the region
$${r\pi\over{r+1}}<\mu<{(r+1)\pi\over{r+2}}\qquad
{\rm with}\quad r=1,2,\cdots,\eqn\coupi$$
$r+2$ string states $S_n$ ($n=1,\cdots,r+2$) are allowed
from Eq.\stringi. The two longest strings
$S_{r+1},S_{r+2}$ represent unbound fermion-antifermion pairs.
The other $r$ strings correspond to the bound states
of a fermion-antifermion pair, i.e. the breathers ($B_n$).
One can solve the integral equations \inteq\ and \inteqi\ using the
Fourier transformation and evaluate the excited energy \exenergy.
{}From this, one can determine
the physical particle spectrum,
$$\eqalign{{\rm\cdot soliton}\ A^{\pm}:\ \
&\quad m_f=m_0{\mu\over{\pi}}{\tan\pi^2/2\mu\over{\pi-2\mu}}
e^{\Lambda(1-\pi/2\mu)}\cr
{\rm\cdot breather}\ B_n:
&\quad m_n=2m_f\sin\left[{n\pi\over{2}}\left({\pi\over{\mu}}
-1\right)\right]\quad n=1,\cdots,r.\cr}\eqn\massii$$
This is the mass spectrum derived in the semi-classical
approximation in \massi.

The $S$-matrices of the physical particles (the string states $S_n$)
can be derived
as the products of the scattering amplitudes of pseudo-particles
which constitute the string states,
$$\eqalign{S^n_m(\t)&=\exp\left[\Phi^n_m(\t)\right]
\qquad{\rm for}\quad n,m=1,\cdots,r+2\cr
&=\prod_{j=0}^{n-1}\prod_{k=0}^{m-1}\exp\left[\phi\left(\t
+i(\pi-\mu)(2j-2k+m-n)\right)\right].\cr}\eqn\besma$$
These results give the exactly same $S$-matrices
as Eqs.\sgsma, \sgsma\ and \sgmn.\refmark\Korepini\

The SG theory in the repulsive regime has been solved by
V. Korepin.\refmark\Korepin\
The essecial difference in this regime arises from the fact that
the allowed strings ($S_n$),
$$\eqalign{\t_l&=\alpha+i\pi+i\mu(n-1-2l),\qquad{\rm for}\quad
0<\mu<{\pi\over{2}},\cr
l&=0,1,\cdots,n-1,\qquad{\rm with}\quad
\mu(n-1)<\pi,\cr}\eqn\stringii$$
have negative energies from Eq.\energy,
$$E\left[S_n\right]=-m_0{\sin(n\mu)\over{\sin\mu}}\cos\alpha,\eqn\enerii$$
with respect to the vacuum energy of a pseudo-particle ($E_0=-m_0\cos\alpha$)
in the previous Dirac sea. This means one needs to construct new vacuum
(Dirac sea) where the negative energy states created by the strings are
completely filled.

In the region of the repulsive regime
$${\pi\over{r+2}}<\mu<{\pi\over{r+1}}\quad r=1,2,\cdots,\eqn\coupii$$
there are $r$ string ($S_n(\alpha)$ $n=1,\cdots,r$) states which should be
filled.
\foot{
In fact, $r+2$ strings $S_n$ are allowed from Eq.\stringii.
But one can show using Eq.\enerii\ that the energy of $S_{r+2}$
is positive while that of $S_{r+2}$ is positive relative to the energy
of a pseudo-particle.}
In other words,
the true vacuum is a $r$-component condensate in which all permitted rapidity
states $\alpha$ of
these composite particles are completely filled.
The densities $\rho_n(\t)$ of these strings $S_n$ in the new Dirac sea can be
determined as before by the PBC in the continuum limit
$$2\pi\rho_n(\alpha)=m_0{\sin(n\mu)\over{\sin\mu}}
\cosh\alpha-\sum_{m=1}^{r}\int_{-\Lambda}^{\Lambda}
d\alpha^{\prime}\left[{\partial\over{\partial\alpha}}\Phi^n_m
(\alpha-\alpha^{\prime})\right]\rho_m
(\alpha^{\prime}),\eqn\inteqii$$
where the phase $\Phi^n_m$ of the scattering amplitude between two
strings $S_n$ and $S_m$ is equal to
$$\Phi^n_m(\t)=\sum_{j=0}^{n-1}\sum_{k=0}^{m-1}\phi\left(\t
+i\mu(2j-2k+m-n)\right).\eqn\phasei$$

The physical excited states are created by removing strings $S_n$ out of
vacuum (or creating a `hole string'). If a string is removed, the vacuum
is realigned and the densities of the strings in the new Dirac sea
satisfy new integral equation
$$\eqalign{2\pi\rhoh_n(\alpha)=&m_0{\sin(n\mu)\over{\sin\mu}}\cosh\alpha
+\Phi^n_h(\alpha-{\overline{\alpha}})\cr
&-\sum_{m=1}^r\int_{-\Lambda}^{\Lambda}
d\alpha^{\prime}\left[{\partial\over{\partial\alpha}}\Phi^n_m(\alpha
-\alpha^{\prime})\right]\rhoh_m
(\alpha^{\prime}).\cr}\eqn\inteqiii$$
The excited energies can be expressed in terms of these densities as
follows:
$$\eqalign{E_n&=M_n\cosh\t\cr
&=m_0{\sin(n\mu)\over{\sin\mu}}\cosh{\overline{\alpha}}
-m_0\sum_{m=1}^r\int_{-\Lambda}^{\Lambda}d\alpha^{\prime}
\left[\rho_m(\alpha^{\prime})-\rhoh(\alpha^{\prime})\right]\cosh
\alpha^{\prime}.\cr}\eqn\exenergyi$$
The string state $S_r$ is interpreted as the fermion (the SG soliton).
The other string states $S_n$ ($n=1,\cdots,r-1$) correspond to new
particles, called `neutral particles' ($N_n$).
One can find the mass spectrum from Eq.\exenergyi,\refmark\Korepin\
$$\eqalign{
{\rm\cdot soliton}\ A^{\pm}:\ &\quad m_f=M\left[{\sin(r-1)\mu\over{\sin\mu}}+
{\sin(r\mu)\over{\sin\mu}}\tan{\pi\over{2}}\left({\pi\over{\mu}}
	-r-1\right)\right]\cr
{\rm\cdot neutrals}\ N_n:&\quad
M_n=2M{\sin(n\mu)\over{\tan\mu}}
		\quad n=1,\cdots,r-1,\cr
{\rm with}&\quad\ M=m_0{\exp[(2\mu-\pi)\Lambda/2\mu]\over{\pi-2\mu}}
\quad \t={\pi{\overline{\alpha}}\over{2\mu}}.
\cr}\eqn\massiii$$

The scattering amplitudes of these physical particles can be computed
from the products of those of each constituent pseudo-particles
like Eq.\phasei.
The $S$-matrices are given as follows:\refmark\Korepin\
$$S_{\rm rep}^{SG}\left(\t,{\g\over{8\pi}}
\right)=S^{SG}\left(\t,{\g_{\rm eff}\over{8\pi}}\right),\eqn\korsmai$$
for the SG soliton and anti-soliton (fermion and antifermion),
$$S^{(n)}_{\rm rep}(\t)=\left({i\exp(\t)+
1\over{\exp(\t)+i}}\right)^{\delta^n_{r-1}},\eqn\korsmaii$$
for $A^{\pm}+N_n\to N_n + A^{\pm}$ ($n=1,2,\cdots,r-1$), and
$$S^{(n,m)}_{\rm rep}(\t)=\left({i\exp(\t)+
1\over{\exp(\t)+i}}\right)^{\delta^m_{n-1}},\eqn\korsmaiii$$
for $N_n +N_m\to N_m +N_n$ ($n,m=1,2,\cdots,r-1$).
Note that the soliton $S$-matrix is the same as the Zamolodchikov's
$S^{SG}(\t,\g/8\pi)$ \sgsma\ with a renormalized
coupling constant $\g_{\rm eff}/8\pi$.
For the coupling constant in the region \coupii,
the renormalized coupling constant is equal to
$${\g_{\rm eff}\over{8\pi}}={\g\over{8\pi}}-(r-1),\quad{\rm and}
\quad 1<{\g_{\rm eff}\over{8\pi}}<2.\eqn\gameff$$
The neutral particles $N_n$ are not the bound states of the fermion and
anti-fermion because the soliton $S$-matrix in Eq.\korsmai\ has no poles
in the physical strip. These are new excitation modes appearing only in
the repulsive regime.
In the region $\pi/3<\mu<\pi/2$ ($r=1$), since $\g_{\rm eff}=\g$,
the soliton $S$-matrix is equal to Eq.\sgsma\ without any additional
particles. Therefore, the Zamolodchikov's results can be
analytically continued to a part of the repulsive regime.

We conclude this section with remarks  on the first-order phase transition
points in the repulsive regime, $\mu=\pi/(r+2)$.
For these values, the Bethe ansatz analysis explained in sect.2 does not
work. The reason is as follows:
{}From Eq.\stringii, there are $r$ strings $S_n$.
(The other two strings $S_{r+2}$ and $S_{r+1}$ have positive and zero
energies relative to that of a pseudo-particle.)
The masses of these strings are degenerate, i.e. $S_n$ has the same mass
as $S_{N-n}$ from Eq.\enerii. If there are several different
particles with same masses, the $S$-matrices become non-diagonal, for
which the Bethe ansatz analysis (the `high-level Bethe ansatz')
becomes very complicated. It has been claimed that
these are the first-order phase transitions.
\refmark{\BKI,\JNW}\

Taking the limit $\mu\to \pi/(r+2)$ of Eq.\massiii, one can find the fermion
(the SG soliton) mass becomes infinite. This would mean
that one should have only neutral particles in the spectrum after removing
the infinite mass solitons.\refmark\IaM\
On the other hand, however, one can choose
an appropriate cut-off parameter in the mass
renormalization such that the SG soliton mass remains finite.
This regularizaion makes the masses of the neutral particles vanish.
This seems to be more consistent with
the results based on the lattice SG theory in which the fermion gets
well-defined mass.\refmark{\BKI,\JNW}\
After removing the massless sector from the massive theory, one obtains
the SG theory with only solitons and anti-solitons in the spectrum.
One can determine the soliton $S$-matrix from the Yang-Baxter equation
which should be the same form as Eq.\sgsma.
Therefore, it seems possible to extend the Zamolodchikov's results to the
special first-order phase transition points in the repulsive regime.
This $S$-matrix has been used to construct the RSG theory which is
connected to the perturbed unitary CFTs.\refmark{\RSG,\ABL,\Ahn}\
{}From now on, we will consider only the unambiguous region \coupii\ in
the repulsive regime.

\endpage

\chapter{Restricted sine-Gordon Theory}

After reviewing the RSG theory based on the Zamolodchikov's results,
we construct the RSG theory in the repulsive regime based on the
Korepin's $S$-matrices  by restricting
the multi-soliton Hilbert space of the SG theory using the
underlying quantum group symmetry.  We connect this RSG theory with the
perturbed minimal CFT ${\cal M}_{p/p+2}$.

\section{The quantum group symmetry of the SG theory}

The quantum group $SL_q(2)$ is defined by the universal enveloping algebra
${\cal U}_q[sl(2)]$ with the commuation relations\refmark\KaR\
$$[J_+,J_-]={q^h-q^{-h}\over{q-q^{-1}}},\qquad
[h,J_{\pm}]=\pm 2 J_{\pm}.\eqn\comm$$
If the deformation parameter $q$ goes to $1$, Eq.\comm\ reduces to the
ordinary $sl(2)$ commutation relations and the quantum
group $SL_q(2)$ to the ordinary $SL(2)$ group.
The ${\cal U}_q[sl(2)]$ forms the Hopf algebra with the
`comultiplication' $\Delta_q$, defined by
$$\eqalign{
\Delta_q(h)&=1\otimes h + h\otimes 1\cr
\Delta_q(J_{\pm})&=q^{h/2}\otimes J_{\pm} + J_{\pm}\otimes q^{-h/2}.\cr
						}\eqn\comul$$
The irreducible representations of $SL_q(2)$ are generated by the
comultiplication $\Delta_q$ which defines tensor product
representations.
Again, Eq.\comul\ reduces to the usual rule of the addition of
angular momentum as $q\to 1$.

The representations of the $SL_q(2)$ are well-defined; there is
one-to-one correspondence to the representation of the ordinary
$SL(2)$ which is represented by half-integer spin $j$. Starting with the
fundamental representation and using Eq.\comul, one can
generate all the irreducible representations with higher spins from the
relation
$$|J,M;j_1,j_2\rangle=
\sum_{m_1,m_2}\left[{j_1\ j_2\ J\atop{m_1 m_2 M}}\right]_q
|j_1,m_1\rangle\otimes|j_1,m_1\rangle,\eqn\decomp$$
with the quantum group analogue of the Clebsch-Gordan (CG) coefficients.
These quantum CG coefficients (and quantum $6j$ (the Wigner-Racah)
symbols) are expressed with `$q$-numbers'\refmark\KaR\
$$[n]={q^n-q^{-n}\over{q-q^{-1}}}, \quad{\rm and}\quad
[n]\to n \quad{\rm as }\quad q\to 1.\eqn\qnum$$

If the deformation parameter $q$ is a a root of unity, one can see from
Eq.\qnum\ that some $q$-CG coefficients (and $q$-$6j$ symbols) become
singular.
For this case, the sensible representation theory of the $SL_q(2)$
is possible by restricting the allowed spins to
$\{ 0,1/2,\cdots,j_{\rm max}\}$.
The $j_{\rm max}$ is determined by the condition that $[2j_{\rm
max}+1]_q=0$, which gives
$$j_{\rm max}={N\over{2}}-1\qquad{\rm for }\quad q^N=\pm 1.\eqn\jmax$$
This restriction on the allowed representations of the $SL_q(2)$ with
$q$ a root of unity leads to the truncation of the multi-soliton Hilbert
space of the SG theory.

The quantum group $SL_q(2)$ can be realized by
the $R$-matrix defined by
$$R(q)(g\otimes 1)(1\otimes g)=(1\otimes g)(g\otimes 1)R(q),\quad
{\rm for}\quad g\in SL_q(2).\eqn\rma$$
With Eq.\comul, this means $[R(q),\Delta_q(g)]=0$ for any $g\in
SL_q(2)$.
For the fundamental representation (spin-$1/2$), $g$ is $2\times 2$
matrix with $q$-determinant 1 ($g_{11}g_{22}-qg_{12}g_{21}=1$) and
the $R$-matrix is given by
$$R(q)=\left(\matrix{q&0&0&0\cr
					 0&1&q-q^{-1}&0\cr
                     0&0&1&0\cr
                     0&0&0&q\cr}\right).\eqn\rmatrix$$
The solution of Yang-Baxter equation can be expressed in terms of the
$R$-matrix as follows:
$${\widehat R}(x,q)=x{\widehat R}(q)-x^{-1}{\widehat R}^{-1}(q)
\quad{\rm with}\quad {\widehat R}={\cal P}R,\eqn\rmatrixi$$
where the permuation ${\cal P}$ is defined by
${\cal P}(V_1\otimes V_2)=V_2\otimes V_1$.

The quantum group $SL_q(2)$ symmetry of the SG theory was discovered by
simply noticing that the SG soliton $S$-matrix can be expressed in terms of
the $R(x,q)$, which satisfies the Yang-Baxter equation.\refmark{\RSG-\RaS}\
The Zamolodchikov's soliton $S$-matrix \sgsma\ can be written as
$$\eqalign{&S^{SG}\left(\theta,{\gamma\over{8\pi}}\right)=
{U(\theta)\over{2\pi i}} R
\left(x=e^{\theta},q\right),\cr
&q=-\exp\left(-{i\pi\over{P}}\right)\quad
{\rm with} \quad P={\gamma\over{8\pi}},\cr}\eqn\sgsmai$$
upto a global gauge transformation.
This result is rather expected because the Zamolodchikov's $S$-matrix
was derived from the Yang-Baxter equation upto an
overall factor.\refmark\Zamol\

Eq.\sgsmai\ means the SG theory has the underlying quantum group symmetry
$SL_q(2)$ where the deformation parameter $q$ is related to the coupling
constant. The soliton and anti-soliton pair forms the fundamental
(spin-$1/2$) representation from the fact that $[S^{SG},\Delta_q]=0$.
The multi-soliton states can be represented by the irreducible
representations with higher spins which are generated by tensor products
like Eq.\decomp.
This quantum group symmetry of the SG theory can be derived
directly from the SG Lagrangian.\refmark\BaL\

\section{the Restricted sine-Gordon theory}

\topinsert
\beginpicture
\setcoordinatesystem units <1pt,1pt>
\linethickness 4.0pt
\setlinear
\plot 0 0 300 0 /
\plot 50 50 50 -50 /
\plot 100 50 100 -50 /
\plot 200 50 200 -50 /
\plot 250 50 250 -50 /
{\fourteenpoint
\put {$j_1$} at 25 25
\put {$j_2$} at 75 25
\put {$\cdot$} at 120 25
\put {$\cdot$} at 140 25
\put {$\cdot$} at 160 25
\put {$\cdot$} at 180 25
\put {$j_{N-1}$} at 225 25
\put {$j_N$} at 275 25
\put {$j^{\prime}_1$} at 25 -25
\put {$j^{\prime}_2$} at 75 -25
\put {$\cdot$} at 120 -25
\put {$\cdot$} at 140 -25
\put {$\cdot$} at 160 -25
\put {$\cdot$} at 180 -25
\put {$j^{\prime}_{N-1}$} at 225 -25
\put {$j^{\prime}_N$} at 275 -25
}
\put {$\pm 1/2$} at -20 0
\put {$\pm 1/2$} at 50 60
\put {$\pm 1/2$} at 100 60
\put {$\pm 1/2$} at 200 60
\put {$\pm 1/2$} at 250 60
\endpicture
\vskip 2.5cm
\centerline{\noindent{\bf Figure 1.} \quad{\tenpoint
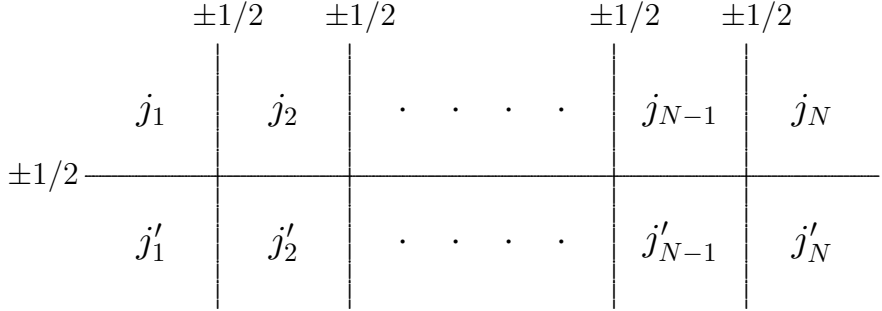
The change of basis from the vertex to the path form}}
\vskip .3cm
\endinsert
As explained in the sect.1, the RSG theory is a modified SG theory
which still preserves the integrability. This is a close
analogy of the lattice RSOS model.
It is well-known that the eight-vertex model is equivalent to the SOS
model, which is obtained simply by changing from the vertex basis to the
path basis.\refmark\ABF\  The SOS model is defined by the spins assigned
on each face. (Fig.1) These two equivalent models are described by $c=1$
CFT at the critical point.
A new integrable lattice model, the RSOS model, is constructed by
restricting the spins upto a maximum spin.
This RSOS model flows to $c<1$ minimal CFTs in the UV limit
(Regime III).

We do exactly the same manipulation for the SG theory to construct
the RSG theory. Since a soliton-antisoliton pair
forms a spin-$1/2$ representation $|1/2,\pm 1/2\rangle$
of $SL_q(2)$, we can decompose
the multi-soliton Hilbert space into the irreducible spaces characterized
by the higher spins (Fig.1):
\def\half{{1\over{2}}}
\def\t{\theta}
$${\cal H}=\sum_{m_i=\pm 1/2}
|\half,m_1\rangle\otimes|\half,m_2\rangle\otimes\cdots
		  |\half,m_N\rangle
  = \sum_{{0\le j_i\le\infty}\atop{|j_i-j_{i+1}|=1/2}}
  |j_1,\cdots,j_N\rangle,\eqn\hilbert$$
with appropriate $q$-CG coefficients.

In this basis, the SG
multi-soliton Hilbert space is spanned by the `kinks' $K_{ab}$ with
$a,b$ as SOS spins satisfying $|a-b|=1/2$ and multi-kink states in Fig.1
$|K_{j_1j_2}\cdots K_{j_{N-1}j_N}\rangle$.
The $S$-matrices for the multi-kink scattering processes can be derived
from the corresponding multi-soliton $S$-matrices with appropriate $q$-CG
coefficients. Like $S$-matrices in the soliton basis, the multi-kink
$S$-matrices can decomposed into the products of two-kink $S$-matrices
which satisfy the Yang-Baxter equation.
One can write down these kink-kink $S$-matrices in terms of the SOS
$R$-matrices which are given by the $q$-$6j$
symbols.\refmark{\RSG-\RaS}\

If $q$ is not a root of unity ($P$ is not a rational),   the SOS SG
theory is well-defined and equivalent to the soliton SG theory.
For $q$ a root of unity (a rational $P$), however, the SOS SG theory
is not well-defined. As we explained before, the allowed irreducible
representations should be restricted to a maximum spin $j_{\rm max}$.
This means the
multi-soliton Hilbert space should be truncated according to the
coupling constant of the SG theory. This is the RSG theory $RSG[P]$
with the kinks $K_{ab}$,
$$\eqalign{
&{\rm For}\ \ P={p\over{q-p}}\quad{\rm with}\quad {\rm coprime\
integers}\ p,\ q\ (q>p),\cr
&K_{ab},\quad{\rm with}\quad 0\le a,b\le j_{\rm max}={p\over{2}}-1\ \ \
|a-b|=\half.\cr}\eqn\kinks$$

The $S$-matrices of the RSG kinks are equal to\refmark{\RSG,\RaS}\ (Fig.2)
$$\eqalign{&S^{RSG}\left[{ab\atop{dc}}\right]
\left(\t,P\right)={U(\t)\over{2\pi
i}}\left({[2a+1][2c+1]\over{
[2d+1][2b+1]}}\right)^{-\t/2\pi i}\ {\cal R}{ab\atop{dc}}(\t) \cr
&{\cal R}{ab\atop{dc}}(\t)=\sinh\left({\t\over{P}}\right)\delta_{db}
\left({[2a+1][2c+1]\over{[2d+1][2b+1]}}\right)^{1/2} + \sinh
\left({i\pi-\t\over{P}}\right)\delta_{ac}, \cr}\eqn\Srsg$$
for the process $\mid K_{da_1}(\t_1)\rangle + \mid K_{a_2b}(\t_2)\rangle
\rightarrow \mid K_{dc_1}(\t_2)\rangle + \mid K_{c_2b}(\t_1)\rangle$
with $a_1=a_2=a$ and $c_1=c_2=c$. If this condition is
not met, the scattering amplitude becomes zero.
The explicit expressions of all non-vanishing amplitudes are given in
Fig.3.
We will discuss the unitarity and crossing symmetry of these
$S$-matrices in the next subsection.
In the attractive regime $P<1$, the RSG $S$-matrix \Srsg\ has poles
corresponding to the bound states. Since the breathers are singlets of
the $SL_q(2)$, the restriction does not change the breather sector.
The $S$-matrices of the breathers in the RSG theory are still given by
Eqs.\sgn\ and \sgmn.
\topinsert
\centerline{
\beginpicture
\setcoordinatesystem units <1pt,1pt>
\fourteenpoint
\linethickness 4.0pt
\arrow <4pt> [.3,.5] from -20 55 to   0 55
\arrow <4pt> [.3,.5] from -20 55 to -20 35
\put {$x$} at 5 55
\put {$t$} at -20 30
\put {$K_{da}(\theta_1)+K_{ab}(\theta_2)\to
K_{dc}(\theta_2)+K_{cb}(\theta_1)$} [lb] at 80 40
\setlinear
\plot 0 25 50 -25 /
\plot 0 -25 50 25 /
\put {$d$} at 2 0
\put {$b$} at 49 0
\put {$a$} at 25 19
\put {$c$} at 25 -18
\put {:\qquad $S^{RSG}\left[{ab\atop{dc}}\right]\left(\t,{\g\over{8\pi}}
\right)\quad (\theta=\theta_1-\theta_2)$}
[lb] at 90 -10
\endpicture
}
\vskip 0.7cm
\centerline{\noindent{\bf Figure 2.} \quad{\tenpoint
The RSOS $S$-matrices of the RSG theory}  }
\vskip 0.3cm
\endinsert

\topinsert
\beginpicture
\setcoordinatesystem units <1pt,1pt>
\linethickness 4.0pt
\setlinear
\plot 0  25 50 -25 /
\plot 0 -25 50  25 /
\put {$(1)$} at -20 0
\put {$l$}  at   2 0
\put {$l\pm 1$} at 49 0
\put {$l\pm{1\over{2}}$} at 25  19
\put {$l\pm{1\over{2}}$} at 25 -18
\setlinear
\plot 0  -50 50 -100 /
\plot 0 -100 50  -50 /
\put {$(2)$} at -20 -75
\put {$l$}  at   2  -75
\put {$l$}  at  49  -75
\put {$l+{1\over{2}}$} at 25  -56
\put {$l+{1\over{2}}$} at 25  -93
\setlinear
\plot 0 -125 50 -175 /
\plot 0 -175 50 -125 /
\put {$(3)$} at -20  -150
\put {$l$}   at   2  -150
\put {$l$}   at  49  -150
\put {$l-{1\over{2}}$} at 25 -131
\put {$l-{1\over{2}}$} at 25 -168
\setlinear
\plot 0 -200 50 -250 /
\plot 0 -250 50 -200 /
\put {$(4)$} at -20 -225
\put {$l$}   at   2 -225
\put {$l$}   at  49 -225
\put {$l\mp{1\over{2}}$} at 25 -206
\put {$l\pm{1\over{2}}$} at 25 -243
{\fourteenpoint
\put {:\quad ${\cal R}(\theta)
=\sinh\left[{i\pi-\theta\over{P}}\right]$} [lb] at 70 -10
\put {:\quad ${\cal R}(\theta)
={\sin[\pi/P]\over{\sin[(2l+1)\pi/P]}}\
\sinh\left[{i\pi(2l+1)+\theta\over{P}}\right]$} [lb] at 70 -85
\put {:\quad ${\cal R}(\theta)
={\sin[\pi/P]\over{\sin[(2l+1)\pi/P]}}\
\sinh\left[{i\pi(2l+1)-\theta\over{P}}\right]$}
[lb] at 70 -160
\put {:\quad ${\cal R}(\theta)
={\sqrt{\sin[2l\pi/P]\sin[(2l+2)\pi/P]}\over{
\sin[(2l+1)\pi/P]}}\
\sinh\left[{\theta\over{P}}\right]$}
[lb] at 70 -235
}
\endpicture
\vskip .7cm
\centerline{\noindent{\bf Figure 3.} \quad{\tenpoint
The RSG $S$-matrices}}
\vskip .3cm
\endinsert
One can identify the RSG theory with the perturbed minimal CFTs under
the conjecture that
the truncation of the multi-soliton Hilbert space due to the quantum
group symmetry is equivalent to adding a background charge to the SG
theory.\refmark{\RSG,\RaS}\
This conjecture can be justified by the argument that the truncation
changes the SG periodic potential to the Landau-Ginzburg potential which
describes the minimal CFTs.
The SG theory in Eq.\sglag\ can be interpreted as the $c=1$ free boson
theory perturbed by the potential,\refmark\EaY\
$\cos\beta\phi=1/2[\exp(i\beta\phi)+\exp(-i\beta\phi)]$.
If one introduces a background charge, one of the perturbation is
identified with the screening operator and the other with the perturbation.
For the background charge $\alpha_0=(p-q)/\sqrt{4pq}$,
($c=1-6(p-q)^2/pq$), if $\Delta[\exp(i\beta\phi)]=1$,
$$\Delta[\exp(-i\beta\phi)]=
{2p-q\over{q}}=\Delta[\Phi_{1,3}].\eqn\misc$$
This leads to the following identification:
$$RSG\left[{\gamma\over{8\pi}}={p\over{q-p}}\right]=\
{\cal M}_{p/q}\quad  + \quad\Phi_{1,3}.\eqn\scheme$$
This connection of the RSG theory to perturbed CFT is rather intuitive.
The consisitency check has been made by
the thermodynamic Bethe ansatz.\refmark{\zamolii,\AaN}\

\section{the unitarity condition on the RSG $S$-matrices}

It is easy to check the
crossing symmetry of Eq.\Srsg\
$$S^{RSG}\left[{ab\atop{dc}}\right]
(\t)=S^{RSG}\left[{bc\atop{ad}}\right](i\pi-\t) \quad{\rm with}\quad
C|K_{ab}\rangle=|K_{ba}\rangle,\eqn\cross$$
where $C$ is the charge conjugation operator for the kinks.
{}From Eq.\kinks\ there are $p-2$ kinks $K_{a,a+1/2}$ and their anti-kinks
$K_{a+1/2,a}$ ($a=0,1/2,\cdots,p/2-1$). A kink and an anti-kink are
identified (Majorana) if they have the same scattering
amplitude.\refmark\Ahn\

The unitarity of the RSG $S$-matrix \Srsg\ gives the condition on the
deformation parameter. The unitarity of a $S$-matrix is
$$S(\theta)^{\dagger}S(\theta)=1\quad\Rightarrow
\quad S(\theta)S^T(-\theta)=1,\eqn\unitarity$$
if $S^{*}(\theta)=S(-\theta)$.
In Eq.\utheta, $U^{*}(\theta)=U(-\theta)$ and
$U(\theta)U(-\theta)\propto 1$. Since all the $q$-numbers are real, the
second prefactor also satisfies this. The $R$-matrices in the RSOS
basis have been known to satisfy\refmark\KaR\
$$\sum_e {\cal R}{ab\atop{de}}(\theta){\cal R}{eb\atop{dc}}(-\theta)=
\delta_{ac}\left[2\cos{2\pi\over{P}}-2\cosh{2\theta\over{P}}
\right].\eqn\runit$$
Combining these together, one can prove the relation
$$\sum_e  S^{RSG}\left[{ab\atop{ed}}\right]
(\theta)S^{RSG}\left[{eb\atop{cd}}\right](-\theta)=
\delta_{ac}.\eqn\sunit$$
Therefore, as far as the relation
${\cal R}{ab\atop{dc}}^{\star}(\theta)=
-{\cal R}{ab\atop{dc}}(-\theta)$ is satisfied,
\foot{Note an extra negative sign which compensates another $-1$ arising
from the complex conjugation of $2\pi i$ in Eq.\Srsg.}
the RSG $S$-matrix is unitary.

{}From the explicit expressions of ${\cal R}{ab\atop{dc}}$ in Fig.3,
this is indeed the case except the last one ($4$) with the square root.
If the factor inside the square root becomes negative, the ${\cal R}$
becomes pure imaginary and gets an $-1$ by the complex
conjugation, which breaks the unitarity.
Therefore, the RSG $S$-matrices can be unitary
only for $P$ which makes the factor inside the square root
positive for all $l=0,1/2,\cdots,p/2-1$.
This is possible only for the following values of
$P(=\g/8\pi)$:\refmark\RaS\
$${\gamma\over{8\pi}}={p\over{q-p}}={N\over{Nk+1}},\quad {\rm for}\ \
N\ge 2,\quad {\rm and}
\quad {3\over{3k+2}}, \quad
{\rm where}\ \ k\ge 0.\eqn\unitP$$
This means that the perturbed minimal CFT ${\cal M}_{p/q}$ can have the
unitary $S$-matrix only for $p,q$ in Eq.\unitP.
This reminds us of the unitarity condition of the minimal CFTs.\refmark\FQS\
In the massive case, the unitary $S$-matrices are possible not only for
the unitary CFTs ${\cal M}_{p/p+1}$ but also for some of non-unitary CFTs.

\section{RSG theory in the repulsive regime}

The previous RSG theory was based on the Zamolochikov's
$S$-matrix in the attractive regime. We derive
the RSG theory in the repulsive regime $1/(r+2)<\mu/\pi<1/(r+1),\
r=1,2,\cdots$. We use a notation ${\overline {RSG}}[P]$
for this new RSG theory.
In this range, the SG $S$-matrices are given by Eqs.\korsmai-\korsmaiii.
The spectrum
is the SG soliton and anti-soliton (with renormalized mass) and $r-1$
neutral particles.
The crucial observation is that the quantum group symmetry $SL_q(2)$ still
exists because the soliton $S$-matrix is the same as the Zamolodchikov's
with the renormalized coupling constant in Eq.\korsmai.
Again, the $(A^+,A^-)$ pair becomes the
spin-$1/2$ representation of the quantum group $SL_q(2)$ from
$$\left[S_{\rm rep}^{SG}\left(\theta,{\gamma\over{8\pi}}\right),
\ \Delta_{q_{\rm eff}}\right]=0,\eqn\newquant$$
with a modified deformation parameter
$$q_{\rm eff}=-\exp\left(-{i\pi\over{P_{\rm eff}}}\right),\quad{\rm with}
\quad P_{\rm eff}={\gamma\over{8\pi}}-(r-1).\eqn\newdef$$

Using this quantum group $SL_{q_{\rm eff}}(2)$ symmetry,
one can restrict the multi-soliton
Hilbert space for a rational value of $P_{\rm eff}$.
Since $1<P_{\rm eff}<2$ from Eq.\gameff, we parametrize a rational
$P_{\rm eff}$ with two coprime integers $n,m$ $(n>m)$
$$P_{\rm eff}=1+{m\over{n}}\quad\Rightarrow\quad
{\g\over{8\pi}}={nr+m\over{n}}.\eqn\para$$
This means schematically the following relationship:
$${\overline {RSG}}\left[{nr+m\over{n}}\right]=
RSG\left[{n+m\over{n}}\right].\eqn\schemei$$
Therefore, the kink $S$-matrices of the new RSG theory ${\overline
{RSG}}[P]$ are given by $S^{RSG}\left(\t,P_{\rm eff}\right)$ in
Eq.\Srsg\ and Fig.3. The neutral particle sector, being singlets of the
quantum group, is unchanged by the restriction. The $S$-matrices of
neutrals with neutrals or kinks are Eqs.\korsmaii\ and \korsmaiii.

{}From the unitarity condition on the RSG $S$-matrices,
$P_{\rm eff}$ should be one of the values given in Eq.\unitP.
Since $P_{\rm eff}>1$, only solution is $P_{\rm eff}=3/2$.
{}From Eq.\para, the only RSG theory with the unitary kink $S$-matrices
in the repulsive regime is ${\overline {RSG}}[p/2]$ ($p=2r+1$).
The connection of this RSG theory to the perturbed CFTs follows exactly
the same way as before, which can be expressed schematically from
Eq.\scheme,
$${\overline {RSG}}\left[{p\over{2}}\right]=\
{\cal M}_{p/p+2}\ +\ \Phi_{1,3},
\quad p=2r+1,\ r=1,2,\cdots.\eqn\schemei$$

\topinsert
\centerline{
\beginpicture
\setcoordinatesystem units <1pt,1pt>
\linethickness 4.0pt
\fourteenpoint
\setlinear
\plot 0  25 50 -25 /
\plot 0 -25 50  25 /
\put {$0$} at  2 0
\put {$0$} at 49 0
\put {${1\over{2}}$} at 25  19
\put {${1\over{2}}$} at 25 -18
\put {$S_{K,{\overline K}}$} at 25 -50
\plot 100  25 150 -25 /
\plot 100 -25 150  25 /
\put {${1\over{2}}$} at 102 0
\put {${1\over{2}}$} at 149 0
\put {$0$} at 125  19
\put {$0$} at 125 -18
\put {$S_{{\overline K},K}$} at 125 -50
\endpicture
}
\vskip 0.5cm
\centerline{\noindent{\bf Figure 4.} \quad{\tenpoint
${\overline{RSG}}[(2r+1)/2]$ kink $S$-matrices}}
\endinsert
For $P_{\rm eff}=3/2$, there are two
kinks $K=K_{0,1/2}$ and ${\overline K}=K_{1/2,0}$ and are
related to each other by the charge conjugation $C(K)={\overline K}$.
The kink $S$-matrices $S^{RSG}(\t,3/2)$ are equal to (Fig.4)
$$S_{K,{\overline K}}(\theta)=
-S_{{\overline K},K}(\theta)=-i\tanh\left[{1\over{2}}
\left(\theta-{i\pi\over{2}}\right)\right].\eqn\newrsgi$$
These two scattering amplitudes, related by the crossing symmetry, are
same upto overall sign.
This makes one to identify two kinks, $K={\overline K}$. In this case,
the crossing relation of Eq.\newrsgi\ gets a factor $-1$.
This is the `$\star$-violated' phenomena.\refmark\RaS\
Other scattering amplitudes (kink-neutral, neutral-neutral)
are given by Eqs.\korsmaii\ and \korsmaiii,
$$\eqalign{
S_{\rm K,N}(\theta)&=\left({i\exp(\t)+1\over{\exp(\t)+i}}\right)
^{\delta^n_{r-1}},\quad n=1,2,\cdots,r-1\cr
S_{\rm N,N}(\theta)&=
\left({i\exp(\t)+1\over{\exp(\t)+i}}\right)^{\delta^m_{n-1}},\quad
n,m=1,2,\cdots,r-1.\cr}\eqn\newrsgii$$
All these scattering amplitudes are diagonal.


\chapter{Thermodynamic Bethe Ansatz for the RSG theory}

In this section, we use thermodynamic Bethe ansatz (TBA)
to confirm the RSG $S$-matrices \newrsgi\ and \newrsgii\ to be those of
the perturbed minimal CFT ${\cal M}_{p/p+2}$.
The TBA\refmark{\TBA-\AaN}\
is basically the same as the Bethe ansatz method except the
fact that particles involved are the
physical particles. We want to find out the wave functions of the
physical particles to compute the energy spectrum.
The ground state energy is connected with the central
charge of the underlying UV CFT by the relation
$$\eqalign{E_0(R)&\sim -{2\pi\over{R}}\left({c\over{12}}-2\Delta_0\right),
\quad{\rm as}\quad R\to 0,\cr
&=-{2\pi\over{R}}\left({1\over{12}}-{1\over{2pq}}\right)\quad
{\rm for}\quad {\cal M}_{p/q},\cr}\eqn\cench$$
where $R$ is the inverse temperature ($R\to
0$ means the UV limit) and $\Delta_0$ is the minimal conformal dimension
allowed in the CFT (for unitary CFT, $\Delta_0=0$).

The integral equation for TBA from the PBC is
$$2\pi\rho(\t)=-m\cosh\t-\int_{-\infty}^{\infty}
d\t^{\prime}\left[{\partial\over{\partial\t}}\Phi(\t-\t^{\prime})\right]\rho_1
(\t^{\prime}),\eqn\inteqiii$$
where $\rho$ is the density of the allowed states and $\rho_1$ is the
density of the occupied states by the physical particles.
Note that in TBA the density of the
allowed states is not necessarily the same as that of the occupied states.
$\Phi$ is the phase of the physcal scattering amplitude
$$\Phi(\t)=-i\ln S(\t),\eqn\eq$$
if the $S$-matrices are diagonal.
For non-diagonal $S$-matrices, one should diagonalize the transfer
matrix. This has been succeeded only for part of the RSG theories.
\refmark{\zamolii,\AaN}\
Since the $S$-matrices of the new RSG theory are all diagonal, the
diagonal TBA method is enough for our case.

One needs an extra equation to solve the integral equation
involving two unknown
variables like Eq.\inteqiii. In the high
temperature limit, this extra equation comes
from the minimization of the free energy.
This is the TBA equation.
If we introduce the `pseudo-energy' defined by
\def\ep{\epsilon}
$${\rho_{1}\over \rho}={e^{-\ep}\over 1+e^{-\ep}},\eqn\eq$$
the TBA equation is given by
$$Rm\cosh\t=\ep(\t)+{1\over
2\pi}\left[\phi^{\prime}*\ln(1+e^{-\ep})\right](\t),\eqn\tba$$
where the rapidity convolution is
$[\phi^{\prime}*f](\t)=\int_{-\infty}^{\infty}\phi^{\prime}(\t-\t^{\prime})
f(\t^{\prime})d\t^{\prime}$.
The ground state energy is expressed with the pseudo-energy,
$$E_0(R)={m\over{\pi}}\int_0^{\infty}d\t
\cosh\t\ln\left(1+e^{-\ep(\t)}\right).\eqn\ground$$
The TBA equation can be solved in the $R\to 0$ limit and the ground
state energy can be simply expressed in terms of the Rogers dilogarithmic
functions.\refmark{\TBA-\AaN}\

The particle spectrum of the new RSG theory is one kink $K$ ($a=r$) and $(r-1)$
neutrals $N_a$ ($a=1,\cdots,r-1$)
with all different masses where the $S$-matrices $S_{ab}(\t)$ are given
by Eqs.\newrsgi\ and \newrsgii.
Introducing $r$ pseudo-energies $\ep_a(\t)$, we find
the following TBA equations:
$$\eqalign{Rm_a\cosh\t &=\ep_a(\t)+{1\over
2\pi}\sum_{b=1}^r\left[\Phi_{ab}^{\prime}*\ln(1+e^{-\ep_b})\right](\t),\cr
\Phi^{\prime}_{ab}(\t)&=-i{\partial\over{\partial\t}}
\ln S_{ab}(\t)\quad{\rm for}\quad a,b =1,2,\cdots,r.\cr}\eqn\tbai$$
In terms of these pseudo-energies, the ground state energy becomes
$$E_0(R)={m\over{\pi}}\sum_{a=1}^r\int_0^{\infty}d\t
\cosh\t\ln\left(1+e^{-\ep_a(\t)}\right).\eqn\groundi$$

In the UV limit ($R\to 0$), the pseudo-energies $\ep_a(\t)$ have
constant values  $\ep_a(0)$ in the region $|\t|{<\atop{\sim}} 1/R$.
It is straightforward to express the ground state energy $E_0(R)$ as an
integral over the pseudo-energy,\refmark{\TBA-\AaN}\
$$\eqalign{E_0(R)&\sim -{1\over 4\pi R}\sum_{a=1}^r
\int^{\ep_a(\infty)}_{\ep_a(0)}d\ep\left[\ln(1+e^{-\ep})
+ {\ep e^{-\ep}\over 1+e^{-\ep}}\right],\cr
&=-{1\over{\pi R}}\sum_{a=1}^r\left[{\cal
L}\left({x_a\over{1+x_a}}\right)-{\cal
L}\left({y_a\over{1+y_a}}\right)\right],\cr}\eqn\energy$$
where $x_a=\exp[-\ep_a(0)]$ and $y_a=\exp[-\ep_a(\infty)]$.
The Rogers dilogarithmic function ${\cal L}(x)$ is defined by
$${\cal L}(x)=-\half\int^{x}_{0}dt\left[{\ln(1-t)\over t} + {\ln t\over
{(1-t)}}\right].\eqn\dilog$$

\def\half{{1\over{2}}}
The pseudo-energies $\ep_a(0)$ can be
determined by algebraic equations. As $\t\to 0$ in Eq.\tbai,
$x_a$'s satisfy the following algebraic equations:
$$x_a=\prod_{b=1}^r (1+x_b)^{N_{ab}},\quad{\rm for}\quad
a=1,\cdots,r,\eqn\algebra$$
where $N_{ab}=[\Phi_{ab}(\infty)-\Phi_{ab}(-\infty)]/2\pi$
can be easily computed from Eqs.\newrsgi\ and \newrsgii:
$$N = \pmatrix{0&\half&0&\ldots&0&0\cr
					 \half&0&\half&\ldots&0&0\cr
					 0&\half&0&\ldots&0&0\cr
					 \vdots&\vdots&\vdots&\ddots&\vdots&\vdots\cr
					 0&0&0&\ldots&0&\half\cr
					 0&0&0&\ldots&\half&\half\cr}.\eqn\expon$$
We have solved these coupled equations to get the solutions
$$x_a={\sin\left({(a+2)\pi\over{2r+3}}\right)
\sin\left({a\pi\over{2r+3}}\right)\over{\sin^2
\left({\pi\over{2r+3}}\right)}}\quad{\rm for}\quad
a=1,2,\cdots,r.\eqn\solut$$
One can check that these solutions satisfy Eq.\algebra.
In terms of these solutions,
we have found the following sum rule for the Rogers
dilogarithmic functions numerically:
$${1\over{2\pi^2}}\sum_{a=1}^r{\cal L}\left({x_a\over{1+x_a}}\right)
= {r(2r+1)\over{12(2r+3)}}.\eqn\sumi$$

The $\t\to\infty$ limit of Eq.\tbai\ is more subtle because one needs
to know the mass renormalization for the finite temperature. The
mass renormalization will be again given by Eq.\massiii\ with a possible
temperature dependent correction. Without this correction, the
dimensionless combination $RM_a$ in the left-hand side of Eq.\tbai\
becomes $m_0 R\exp[-\Lambda]$ from Eq.\massiii, which vanishes
as $R\to 0$ and $\Lambda\to\infty$.
\foot{In the attractive regime, from Eq.\massii, this term  is equal to
$m_0 R\exp[\Lambda]$, which is non-zero in the double limit.
Therefore, the pseudo-energies $\ep_a(\Lambda)$ in the attractive regime
diverge.}
However, one may get additional finite mass corrections due to the finite
temperature effect. From the dimensional analysis, the correction should
be $\delta M_a\propto 1/R$. We conjecture
that this happens only for the lowest mass
term allowed in the spectrum, that is, $M_1$ of the first neutral
particle $N_1$. This is a natural guess because the lowest mass
particle will be most easily excited by the thermal effect.
This leads to the conjecture
that the left-hand side of
Eq.\tbai\ vanishes for $a=2,3,\cdots,r$ and diverges for $a=1$ as
$\t\to\infty$.
Similar conjecture has been made for TBA computation in ref.[\IaM].

According to this speculation, the algebraic equations for
$y_a=\exp[-\ep_a(\infty)]$ are given by
Eq.\algebra\ with $y_1=0$
$$y_a=\prod_{b=2}^r (1+y_b)^{N_{ab}},\quad{\rm for}\quad
a=2,\cdots,r.\eqn\algebrai$$
The solutions are given by Eq.\solut\ with $r$ replaced by $r-1$,
$$y_a={\sin\left({(a+1)\pi\over{2r+1}}\right)
\sin\left({(a-1)\pi\over{2r+1}}\right)\over{\sin^2
\left({\pi\over{2r+1}}\right)}},\quad{\rm for}\quad
a=1,2,\cdots,r.\eqn\solut$$
Therefore, from the sum rule Eq.\sumi\ with $y_r=0$ one can evaluate the
ground state energy $E_0$ as follows:
$$\eqalign{E_0(R)&\sim -{2\pi\over{R}}\left[{r(2r+1)\over{12(2r+3)}}
-{(r-1)(2r-1)\over{12(2r+1)}}\right],\cr
&=-{2\pi\over{R}}\left[{1\over{12}}-{1\over{2(2r+1)(2r+3)}}
\right].\cr}\eqn\final$$
Comparing this with Eq.\cench, one can confirm that the $RSG[(2r+1)/2]$
theory is indeed the perturbed
minimal CFT ${\cal M}_{2r+1/2r+3}$ by the $\Phi_{1,3}$.

\endpage

\chapter{Discussions}

In this paper, we studied the SG (MTM) theory
and its restricted version (the RSG theory) in the repulsive regime
and identified the RSG theory with the
pertrubed minimal CFT ${\cal M}_{p/p+2}$ by the least relevent operator.
The complete $S$-matrices, satisfying the unitarity and crossing
symmetry, and the particle spectrum have been derived and checked using
the TBA analysis. We discovered the quantum group $SL_q(2)$ of the SG
theory in the repulsive regime where the deformation parameter is
modified. It would be nice to understand this quantum group
symmetry in the repulsive regime starting with the SG
lagrangian.\refmark\BaL\

The RSG theory is a building block for
a wide class of two-dimensional integrable
QFTs.\refmark{\ABL,\Ahn}\
Many new integrable models have been identified with the general CFTs
perturbed by the massive operators along with exact $S$-matrices.
In particular, the complete
$S$-matrices of the supersymmetric
sine-Gordon (SSG) theory has been derived
by `unrestricting' the restricted SSG theory.\refmark\ABL\
The $S$-matrices of the SSG theory have the factorized form
$S^{SSG}(\t)=S^{RSG}(\t)\otimes S^{SG}(\t)$,
where the first factor is the RSG $S$-matrix for $\g/8\pi=4$ which
commutes with the supersymmetric charge and the second one is the
Zamolodchikov's $S$-matrices. This leads us to suggest the exact
$S$-matrices of the SSG theory in the repulsive regime have the
following form:
$$S_{\rm rep}^{SSG}(\t)=S^{RSG}(\t)\otimes S_{\rm
rep}^{SG}(\t).\eqn\ssgi$$
Since the first factor carries the supersymmetric charge, the particle
spectrum should be not only the solitons, anti-solitons, and neutrals
but also their supersymmetric partners.
Furthermore, by restricting the multi-soliton (SSG) Hilbert space, one
can derive the $S$-matrices of
the perturbed superminimal CFTs $SU(2)_2\otimes SU(2)_L/SU(2)_{L+2}$ ($L$ a
half-integer) by the least relevent operator
to be $S^{RSG}\otimes S_{\rm rep}^{RSG}$
where $S_{\rm rep}^{RSG}$'s are given by Eqs.\newrsgi,\newrsgii.
The extension to the CFTs with fractional supersymmetry is
straightforward by considering higher level ($K>2$) for the first
$SU(2)$. We will report the details elsewhere.

We finish this paper with some open questions.
One is an ambiguity in the first-order phase
transition points which we have discussed at the end of the sect.2;
there exist two different pictures of the SG theory with $\g/8\pi=$
integers in the literature (refs.[\RSG] and [\IaM]).
One may need the high-level Bethe ansatz
analysis to resolve the ambiguity.
Another is about our conjecture for the delicate limit of
the infinite rapidity and $R\to 0$ in the TBA computation.
One needs to understand how the mass renormalization in the Bethe ansatz
approach is affected by the finite temperature effect.
Although we could not give the answer to this,
our TBA computation is still a non-trivial check in that the
complicated coupled integral equations satisfy new sum rule of the
Rogers Dilogarithmic functions and correctly reproduce the central
charge.

\ack

It is a pleasure to thank our colleagues at Cornell for helpful
discussions. This work has been supported in part by the National
Science Foundation.

\refout

\end